\documentclass[12pt,preprint]{aastex}

\slugcomment{}

\shorttitle{shock waves and SPH}
\shortauthors{Nejad-Asghar}

\begin{document}

\title{Simulation of Shock Waves by Smoothed Particle Hydrodynamics}

\author{Mohsen Nejad-Asghar}

\affil{School of Physics, Damghan University of Basic Sciences,
Damghan, Iran} \email{nasghar@dubs.ac.ir}

\begin{abstract}
Isothermal and adiabatic shocks, which are produced from fast
expansion of the gas, is simulated with smoothed particle
hydrodynamics~(SPH). The results are compared with the analytic
solutions. The algorithm of the program is explained and the
package, which is written in Fortran, is presented in the appendix
of this paper. It is possible to change (to complete) the program
for a wide variety of applications ranging from astrophysics to
fluid mechanics.
\end{abstract}

\section{Introduction}

Gas dynamical processes are believed to play an important role in
the evolution of astrophysical systems on all length scales.
Smoothed particle hydrodynamics~(SPH) is a powerful gridless
particle method to solve complex fluid-dynamical problems. SPH
has a number of attractive features such as its low numerical
diffusion in comparison to grid based methods. An adequate
scenario for SPH application is the unbound astrophysical
problems, especially on the shock propagation~(see, e.g., Liu \&
Liu 2003). In this way, the basic principles of the SPH is
written in this paper and the simulation of isothermal and
adiabatic shocks are applied to test the ability of this numerical
simulation to produce known analytic solutions.

The program is written in Fortran and is highly portable. This
package supports only calculations for isothermal and adiabatic
shock waves. It is possible to change (to complete) the program
for a wide variety of applications ranging from astrophysics to
fluid mechanics. The program is written in modular form, in the
hope that it will provide a useful tool. I ask only that:
\begin{itemize}
  \item If you publish results obtained using some parts of this
  program, please consider acknowledging the source of the
  package.
  \item If you discover any errors in the program or
  documentation, please promptly communicate the to the author.
\end{itemize}

\section{Formulation of Shock Waves}

An extremely important problem is the behavior of gases subjected
to compression waves. This happens very often in the cases of
astrophysical interests. For example, a small region of gas
suddenly heated by the liberation of energy will expand into its
surroundings. The surroundings will be pushed and compressed.
Conservation of mass, momentum, and energy across a shock front
is given by the Rankine-Hugoniot conditions~(Dyson \& Williams
1997)
\begin{equation}\label{e:rh1}
\rho_1 v_1=\rho_2 v_2
\end{equation}
\begin{equation}
\rho_1 v_1^2+ K\rho_1^\gamma =\rho_2 v_2^2+ K\rho_2^\gamma
\end{equation}
\begin{equation}\label{e:rh3}
\frac{1}{2}v_1^2 + \frac{\gamma}{\gamma-1} K \rho_1^{\gamma-1}=
\frac{1}{2}v_2^2 + \frac{\gamma}{\gamma-1} K \rho_2^{\gamma-1} +Q
\end{equation}
where the equation of state, $p=K\rho^\gamma$, is used. In
adiabatic case, we have $Q=0$, and for isothermal shocks, we will
set $\gamma=1$.

We would interested to consider the collision of two gas sheets
with velocities $v_0$ in the rest frame of the laboratory. In this
reference frame, the post-shock will be at rest and the pre-shock
velocity is given by $v_1=v_0+v_2$, where $v_2$ is the shock front
velocity. Combining equations (\ref{e:rh1})-(\ref{e:rh3}), we have
\begin{equation}\label{e:v2}
v_2=a_0[-\frac{b}{2}+\sqrt{1+\frac{b^2}{4}+(\gamma-1)
(\frac{M_0^2}{2}-q)}]
\end{equation}
where $a_0\equiv \gamma K\rho_1 ^{\gamma-1}$ is the sound speed,
$M_0\equiv v_0/a_0$ is the Mach number, $b$ and $q$ are defined as
\begin{equation}
b\equiv \frac{3-\gamma}{2}M_0+ \frac{\gamma-1}{M_0}q\quad ;\quad
q\equiv \frac{Q}{a_0^2}.
\end{equation}
Substituting (\ref{e:v2}) into equation (\ref{e:rh1}), density of
the post-shock is given by
\begin{equation}\label{e:den}
\rho_2=\rho_1\{1+\frac{M_0}{[-\frac{b}{2}+\sqrt{1+\frac{b^2}{4}+(\gamma-1)
(\frac{M_0^2}{2}-q)}]}\}.
\end{equation}

\section{SPH Equations}

The smoothed particle hdrodynamics was invented to simulate
nonaxisymmetric phenomena in astrophysics~(Lucy 1977, Gingold \&
Monaghan 1977). In this method, fluid is represented by $N$
discrete but extended/smoothed particles (i.e. Lagrangian sample
points). The particles are overlapping, so that all the physical
quantities involved can be treated as continues functions both in
space and time. Overlapping is represented by the kernel
function, $W_{ab} \equiv W(\textbf{r}_a-\textbf{r}_b,h_{ab})$,
where $h_{ab} \equiv (h_a+h_b)/2$ is the mean smoothing length of
two particles $a$ and $b$. The continuity, momentum and energy
equation of particle $a$ are~(Monaghan 1992)
\begin{equation}
\rho_a=\sum_b m_b W_{ab}
\end{equation}
\begin{equation}
\frac{d\textbf{v}_a}{dt}=-\sum_b m_b (\frac{p_a}{\rho_a}+
\frac{p_b}{\rho_b}+ \Pi_{ab}) \nabla_a W_{ab}
\end{equation}
\begin{equation}
\frac{du_a}{dt}=\frac{1}{2} \sum_b m_b (\frac{p_a}{\rho_a}+
\frac{p_b}{\rho_b}+ \Pi_{ab}) \textbf{v}_{ab} \cdot \nabla_a
W_{ab}
\end{equation}
where $\textbf{v}_{ab}\equiv \textbf{v}_a- \textbf{v}_b$ and
\begin{equation}
\Pi_{ab}=\cases{
       \frac{\alpha v_{sig} \mu_{ab}
       +\beta \mu_{ab}^2}{\bar{\rho}_{ab}}, &
       if $\textbf{v}_{ab}.\textbf{r}_{ab}<0$,\cr
       0 , & otherwise,}
\end{equation}
is the artificial viscosity between particles $a$ and $b$, where
$\bar{\rho}_{ab}= \frac{1}{2}(\rho_a+\rho_b)$ is an average
density, $\alpha\sim 1$ and $\beta\sim 2$ are the artificial
coefficients, and $\mu_{ab}$ is defined as its usual form
\begin{equation}
\mu_{ab}=-\frac{\textbf{v}_{ab}
\cdot\textbf{r}_{ab}}{\bar{h}_{ab}}
\frac{1}{r_{ab}^2/\bar{h}_{ab}^2+\eta^2}
\end{equation}
with $\eta\sim 0.1$ and $\bar{h}_{ab}= \frac{1}{2}(h_a+h_b)$. The
signal velocity, $v_{sig}$, is
\begin{equation}
v_{sig}=\frac{1}{2}(c_a+c_b)
\end{equation}
where $c_a$ and $c_b$ are the sound speed of particles. The SPH
equations are integrated using the smallest time-steps
\begin{equation}
\Delta t_a=C_{cour}MIN[ \frac{h_a}{\mid \textbf{v}_a\mid},
(\frac{h_1}{\mid\textbf{a}_1\mid})^{0.5}, \frac{u_a}{\mid du_a/dt
\mid}, \frac{h_a}{\mid dh_a/dt \mid}, \frac{\rho_a}{\mid
d\rho_a/dt \mid}]
\end{equation}
where $C_{cour}\sim 0.25$ is the Courant number.

\section{Results and Prospects}

The chosen physical scales for length and time are $[l]=3.0 \times
10^{16} m$ and $[t]=3.0 \times 10^{13} s$, respectively, so the
velocity unit is approximately $1km.s^{-1}$. The gravity constant
is set $G= 1$ for which the calculated mass unit is $[m]=4.5
\times 10^{32} kg$. There is considered two equal one dimensional
sheets with extension $x= 0.1 [l]$, which have initial uniform
density and temperature of $\sim 4.5\times 10^8 m^{-3}$ and $\sim
10K$, respectively.

Particles with a positive x-coordinate are given an initial
negative velocity of Mach 5, and those with a negative
x-coordinate are given a Mach 5 velocity in the opposite
direction. In adiabatic shock, with $M_0=5$, the post-shock
density must be $2.9$, which is obtained from analytic solution
(\ref{e:den}) with $Q=0$ and $\gamma=2$. The Results of adiabatic
shock are shown in Fig.~1-3. In isothermal shock, with $M_0=5$,
the post-shock density must be $26.9$, which is obtained from
analytic solution Equ.~(\ref{e:den}) with $\gamma=1$. The Results
of isothermal shock are shown in Fig.~4-5. Algorithm of the
program is shown in Fig.~6.


\clearpage
\begin{figure}
\epsscale{.70} \plotone{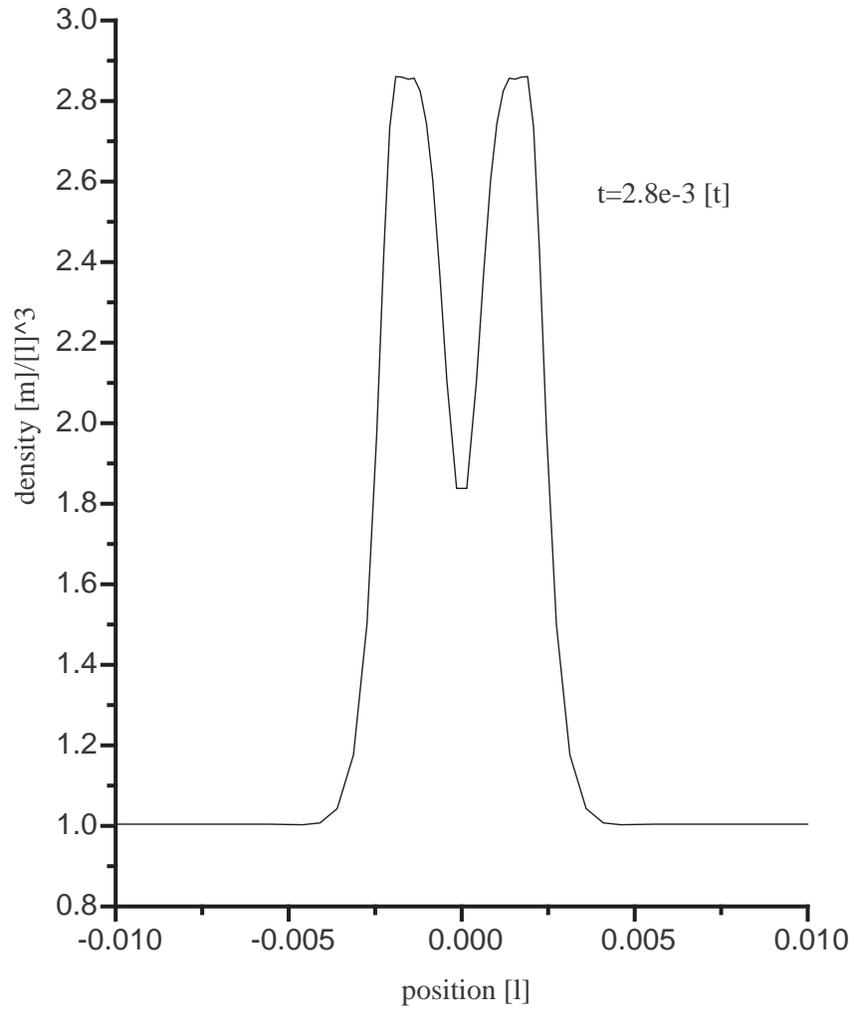} \caption{The density of
adiabatic shock, with $M_0=5$, $Q=0$, and $\gamma=2$.}
\end{figure}
\clearpage
\begin{figure}
\epsscale{.70} \plotone{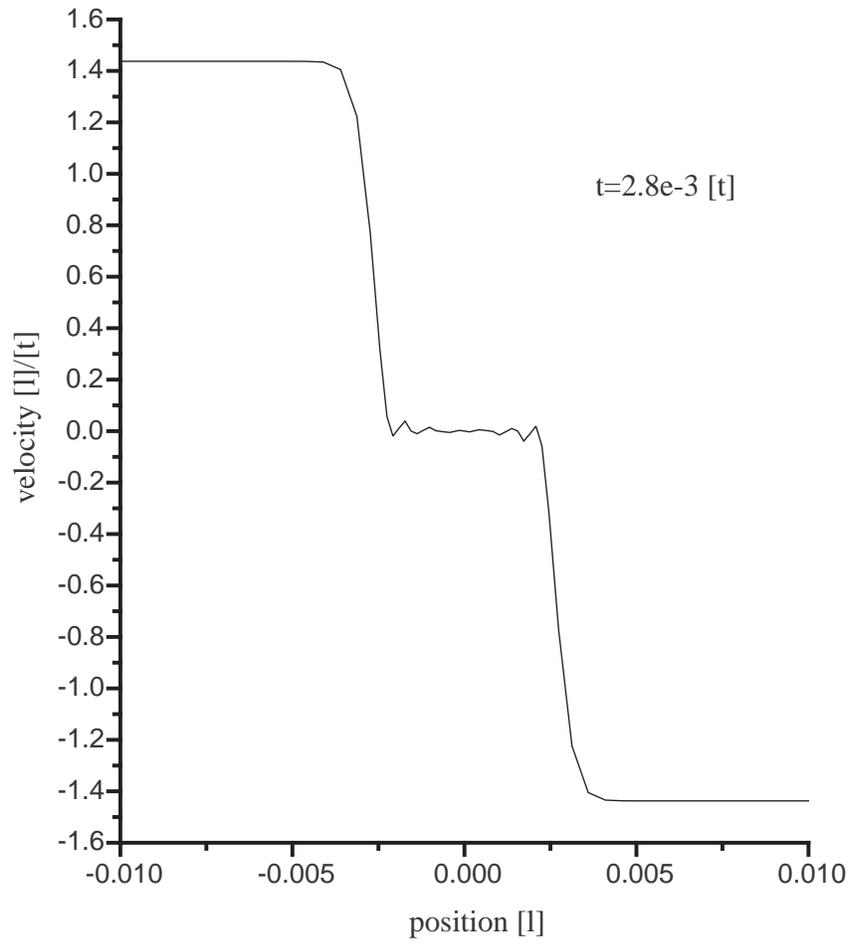} \caption{The velocity of
adiabatic shock, with $M_0=5$, $Q=0$, and $\gamma=2$.}
\end{figure}
\clearpage
\begin{figure}
\epsscale{.70} \plotone{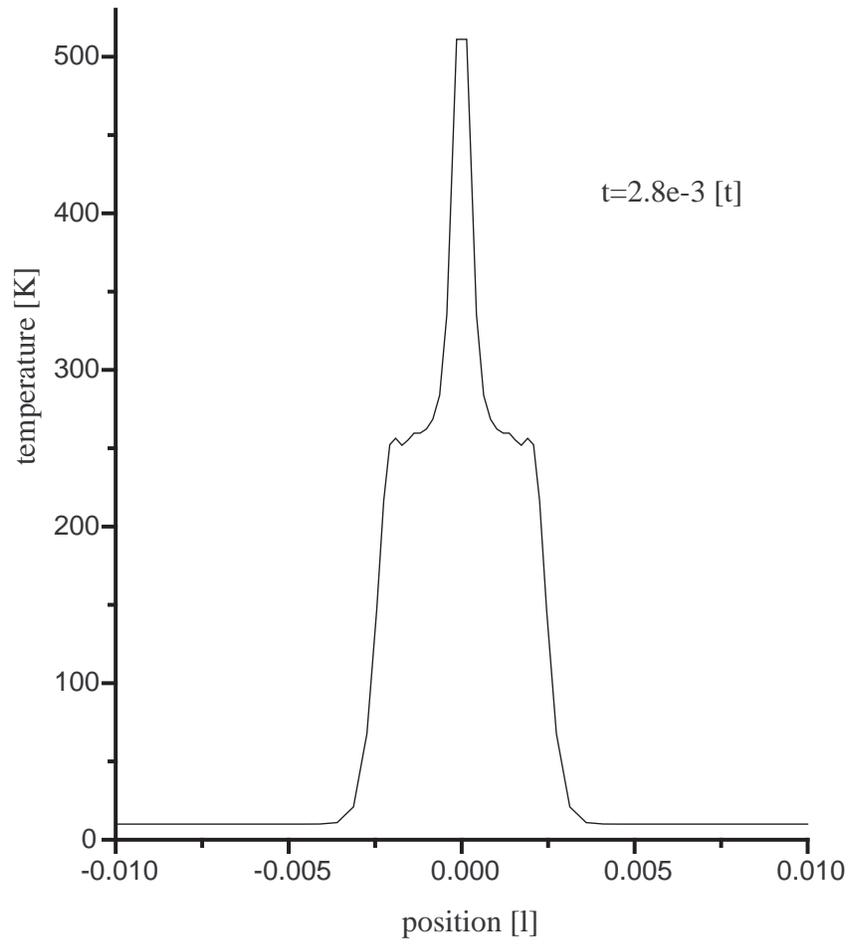} \caption{The temperature of
adiabatic shock, with $M_0=5$, $Q=0$, and $\gamma=2$.}
\end{figure}
\clearpage
\begin{figure}
\epsscale{.70} \plotone{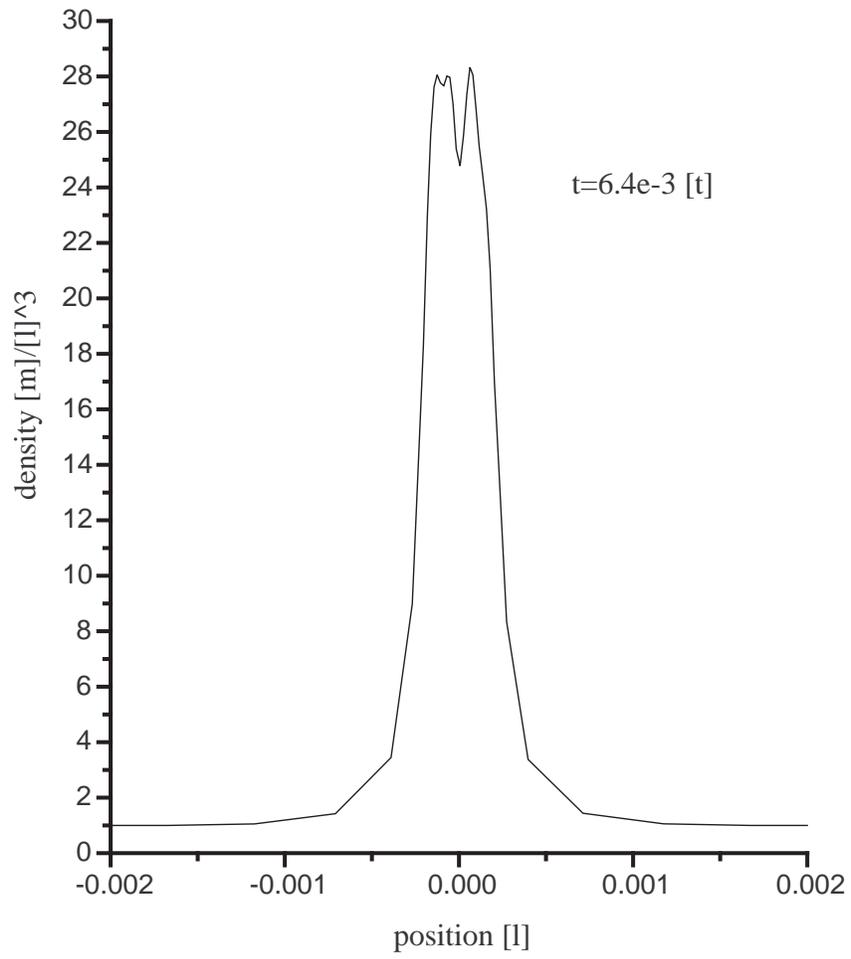} \caption{The density of
isothermal shock, with $M_0=5$ and $\gamma=1$.}
\end{figure}
\clearpage
\begin{figure}
\epsscale{.70} \plotone{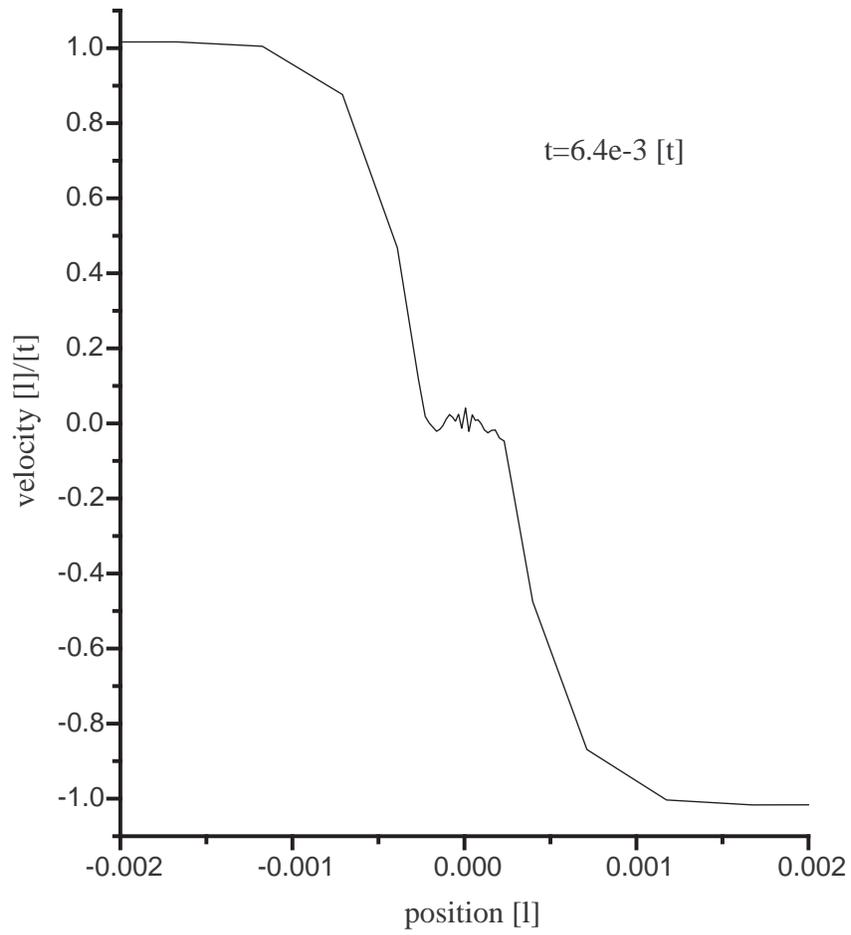} \caption{The velocity of
isothermal shock, with $M_0=5$ and $\gamma=1$.}
\end{figure}
\clearpage
\begin{figure}
\epsscale{.70} \plotone{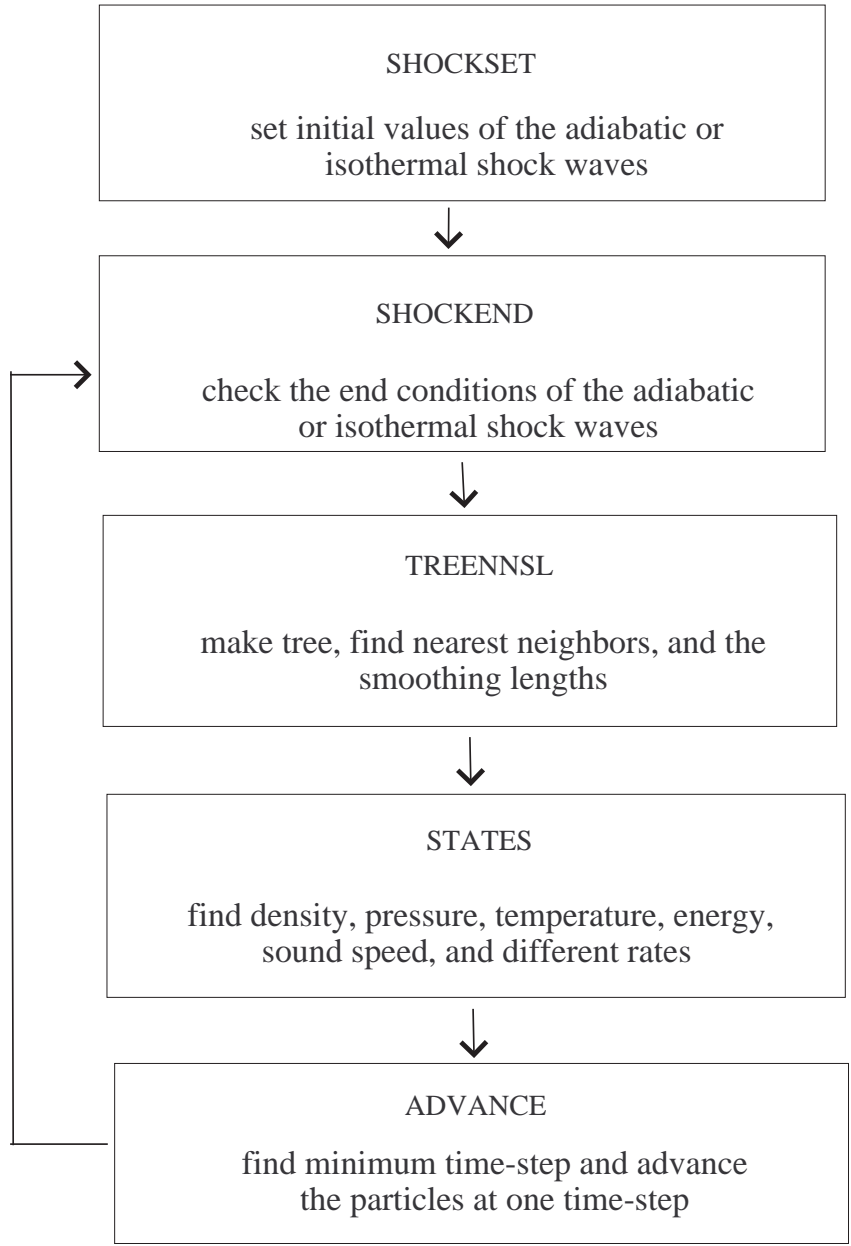} \caption{Algorithm of the
smoothed particle hydrodynamics for simulation of isothermal and
adiabatic shocks.}
\end{figure}
\clearpage
\begin{figure}
\epsscale{0.85} \plotone{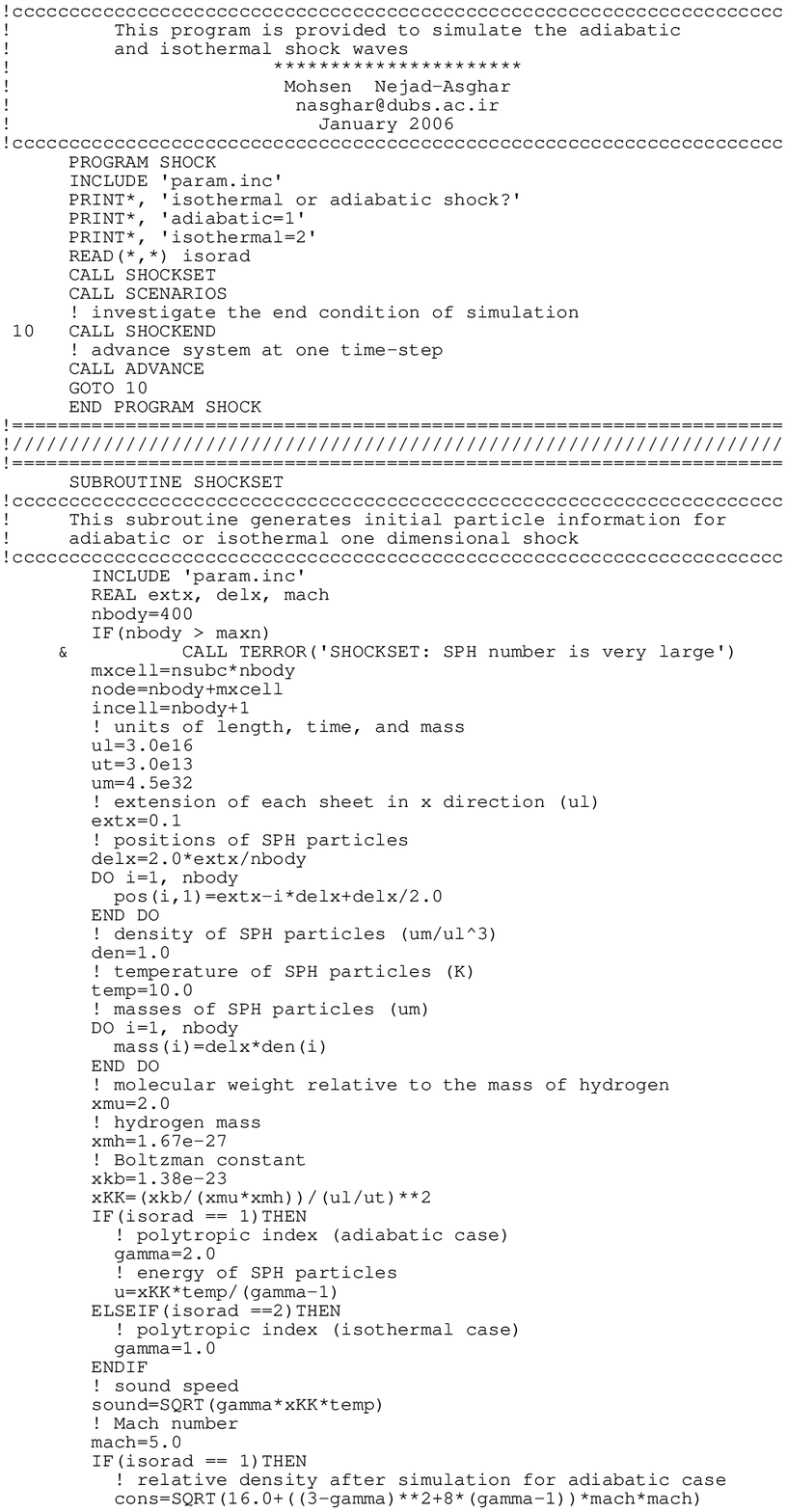}
\end{figure}
\clearpage
\begin{figure}
\epsscale{0.85} \plotone{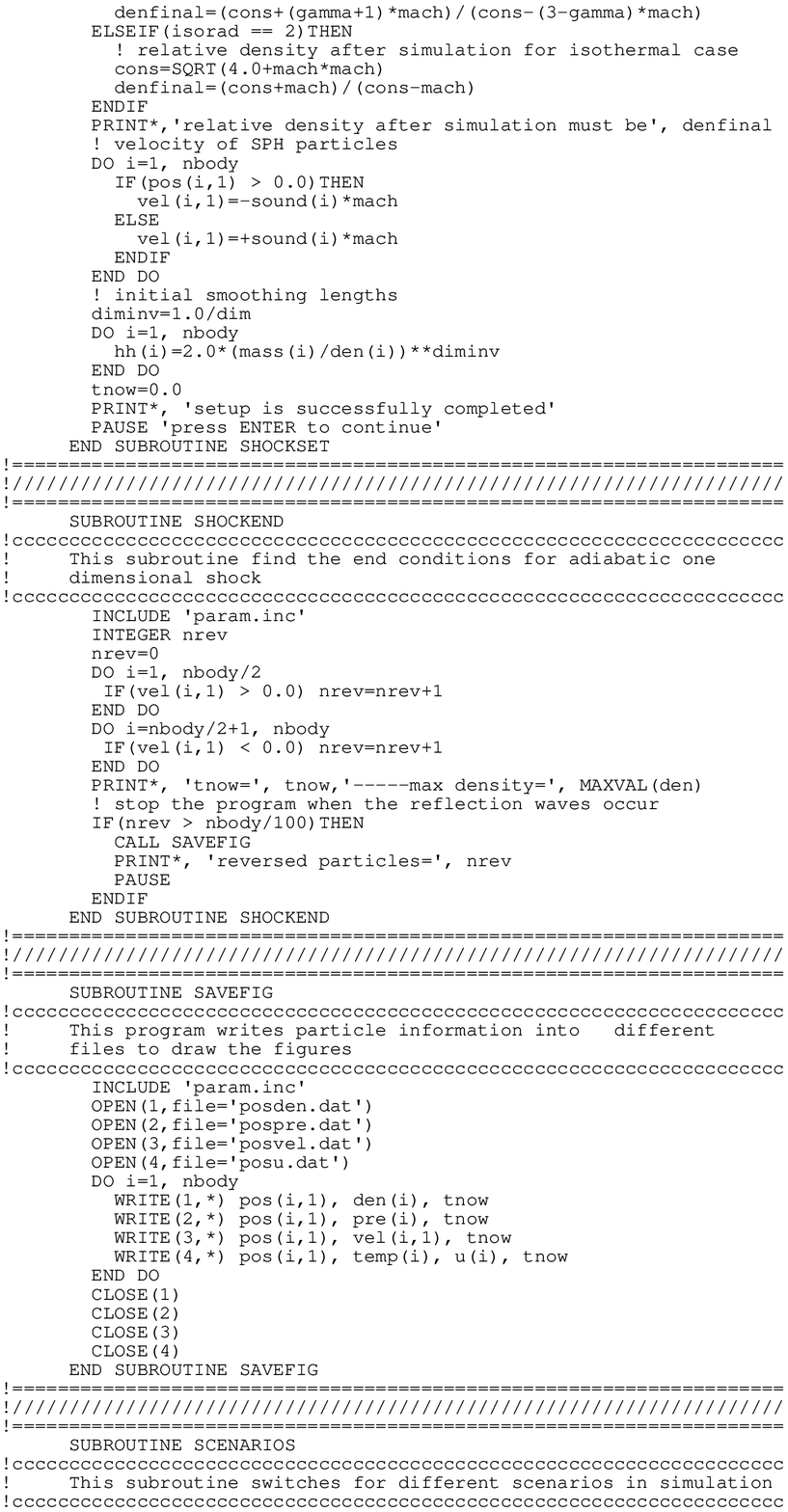}
\end{figure}
\clearpage
\begin{figure}
\epsscale{0.85} \plotone{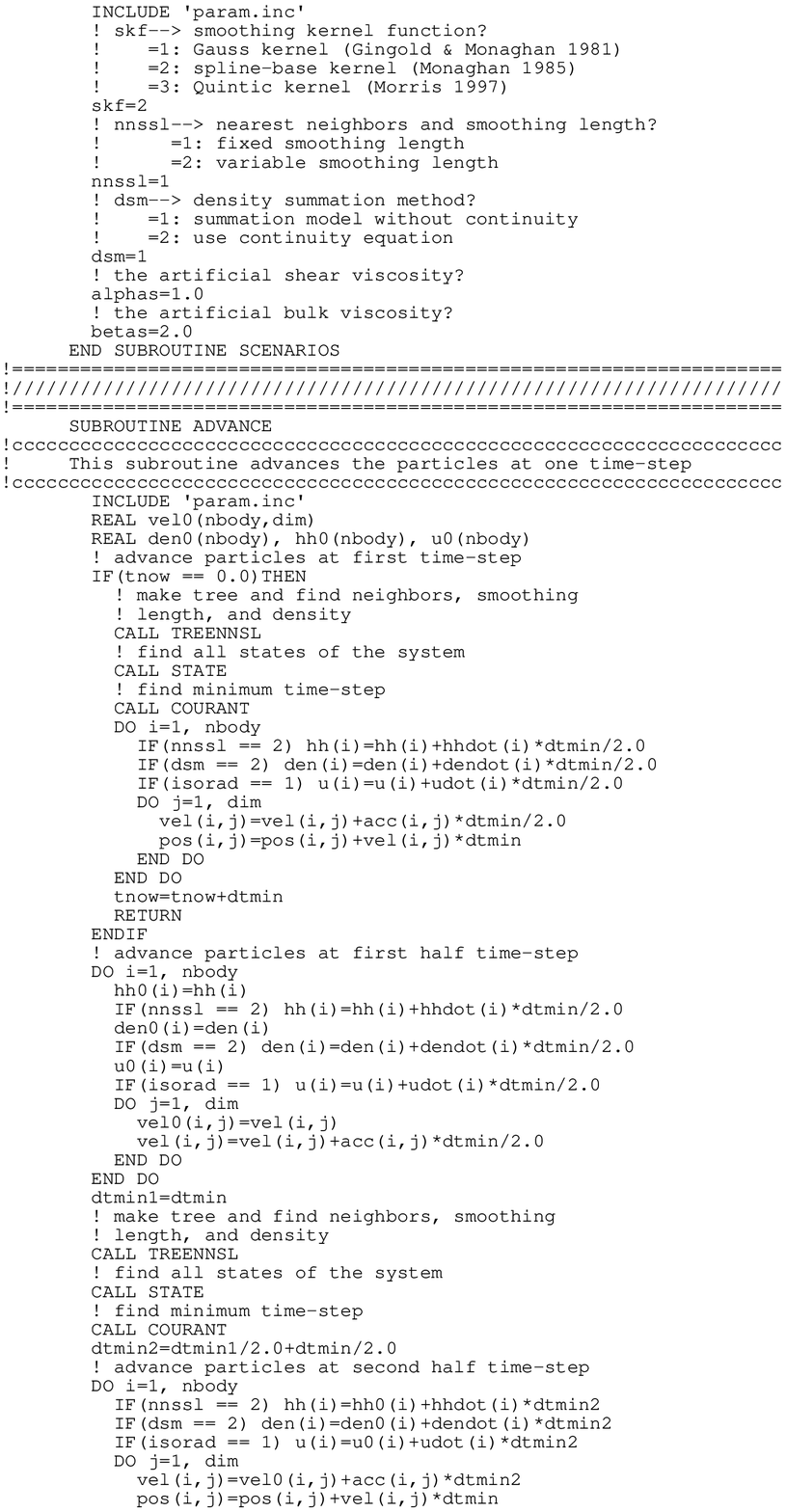}
\end{figure}
\clearpage
\begin{figure}
\epsscale{0.85} \plotone{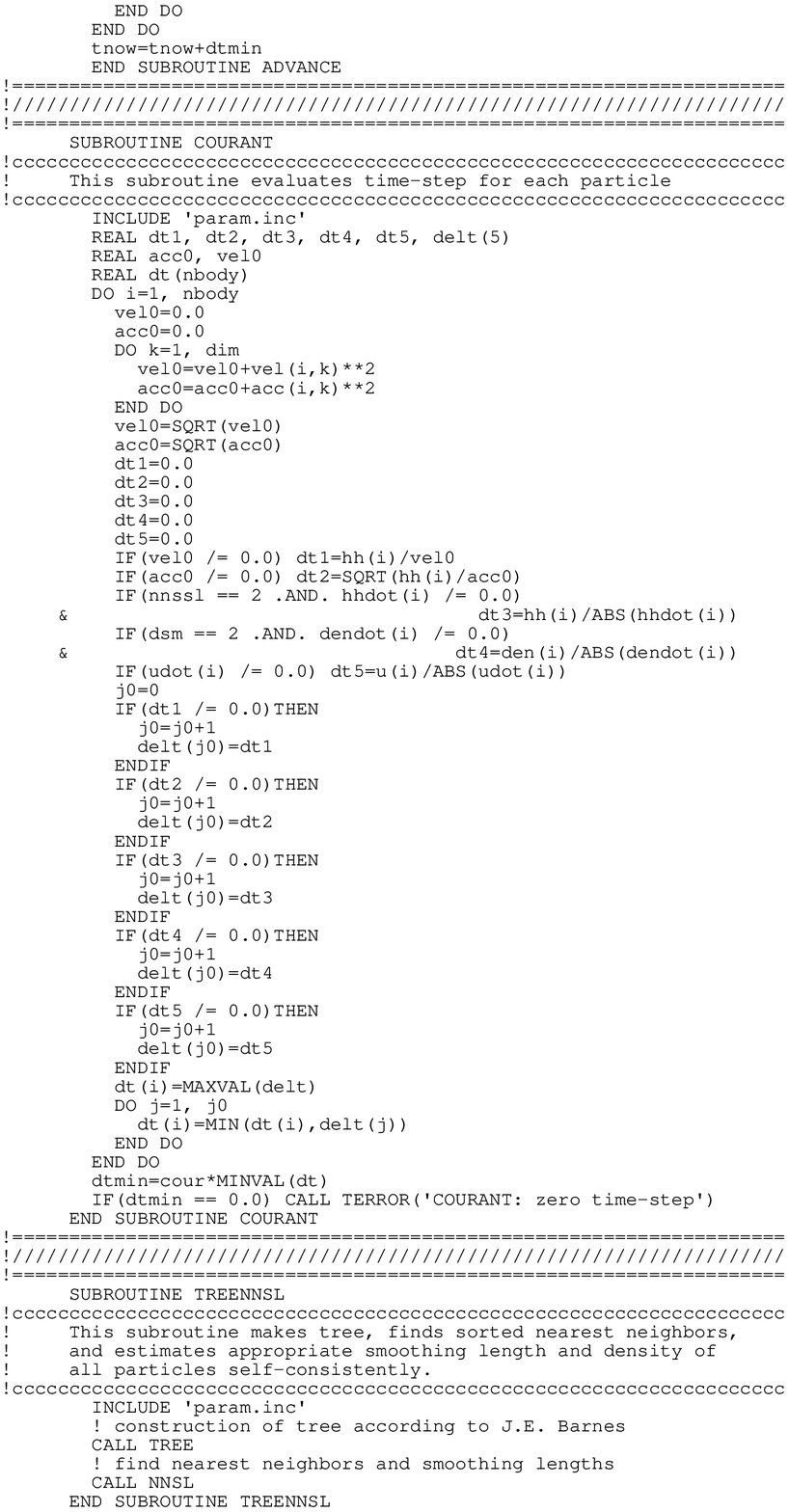}
\end{figure}
\clearpage
\begin{figure}
\epsscale{0.85} \plotone{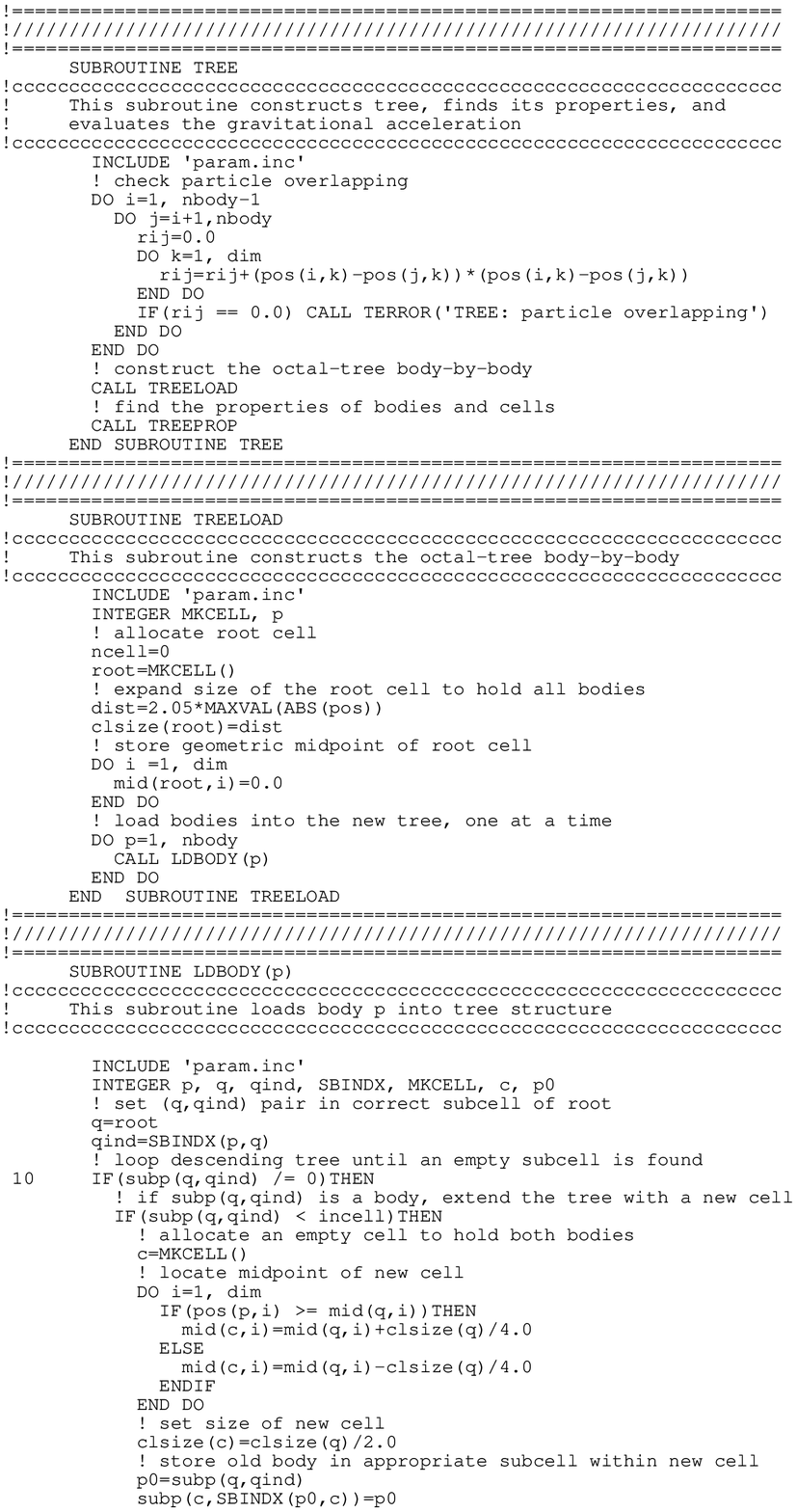}
\end{figure}
\clearpage
\begin{figure}
\epsscale{0.85} \plotone{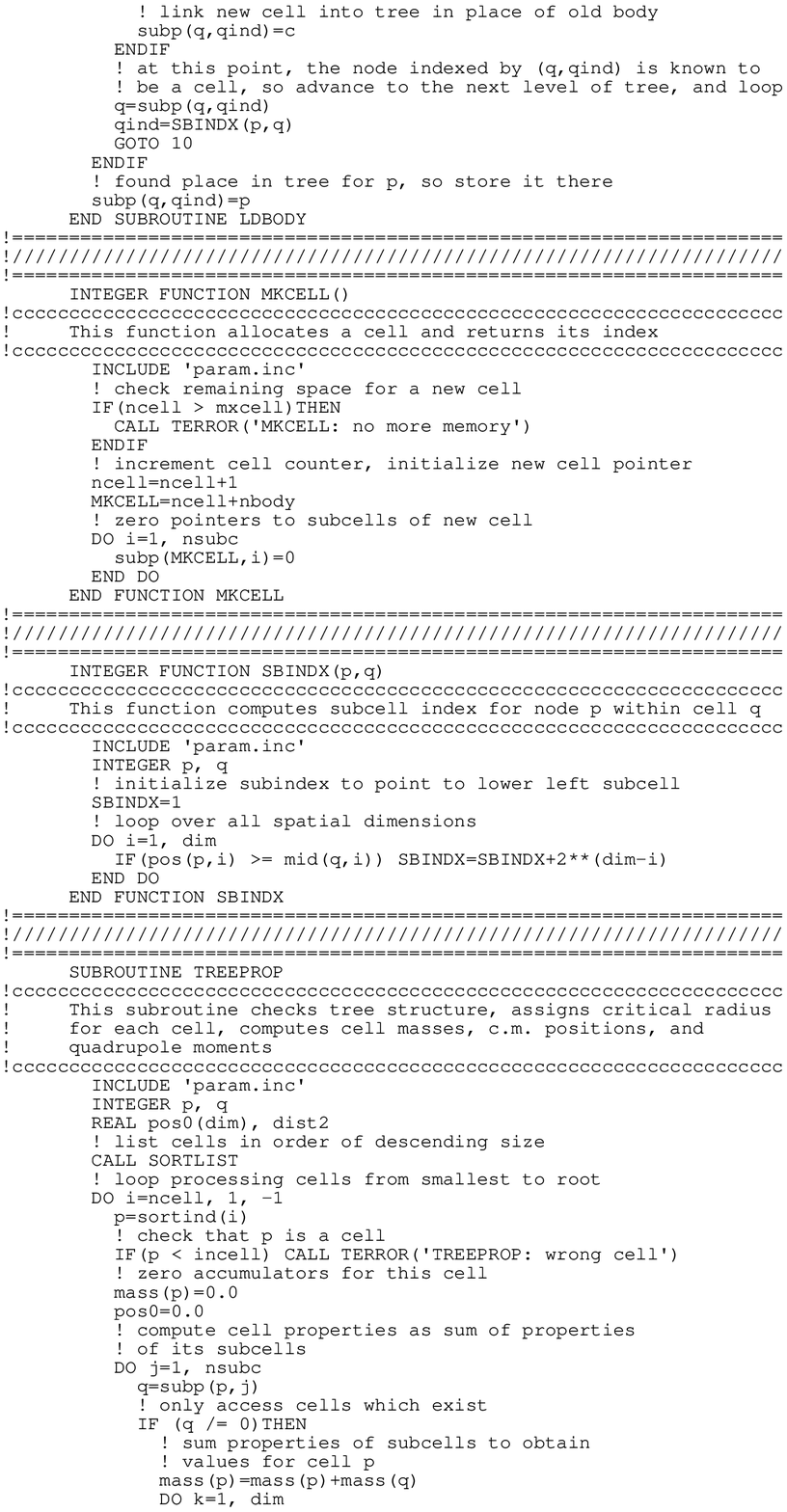}
\end{figure}
\clearpage
\begin{figure}
\epsscale{0.85} \plotone{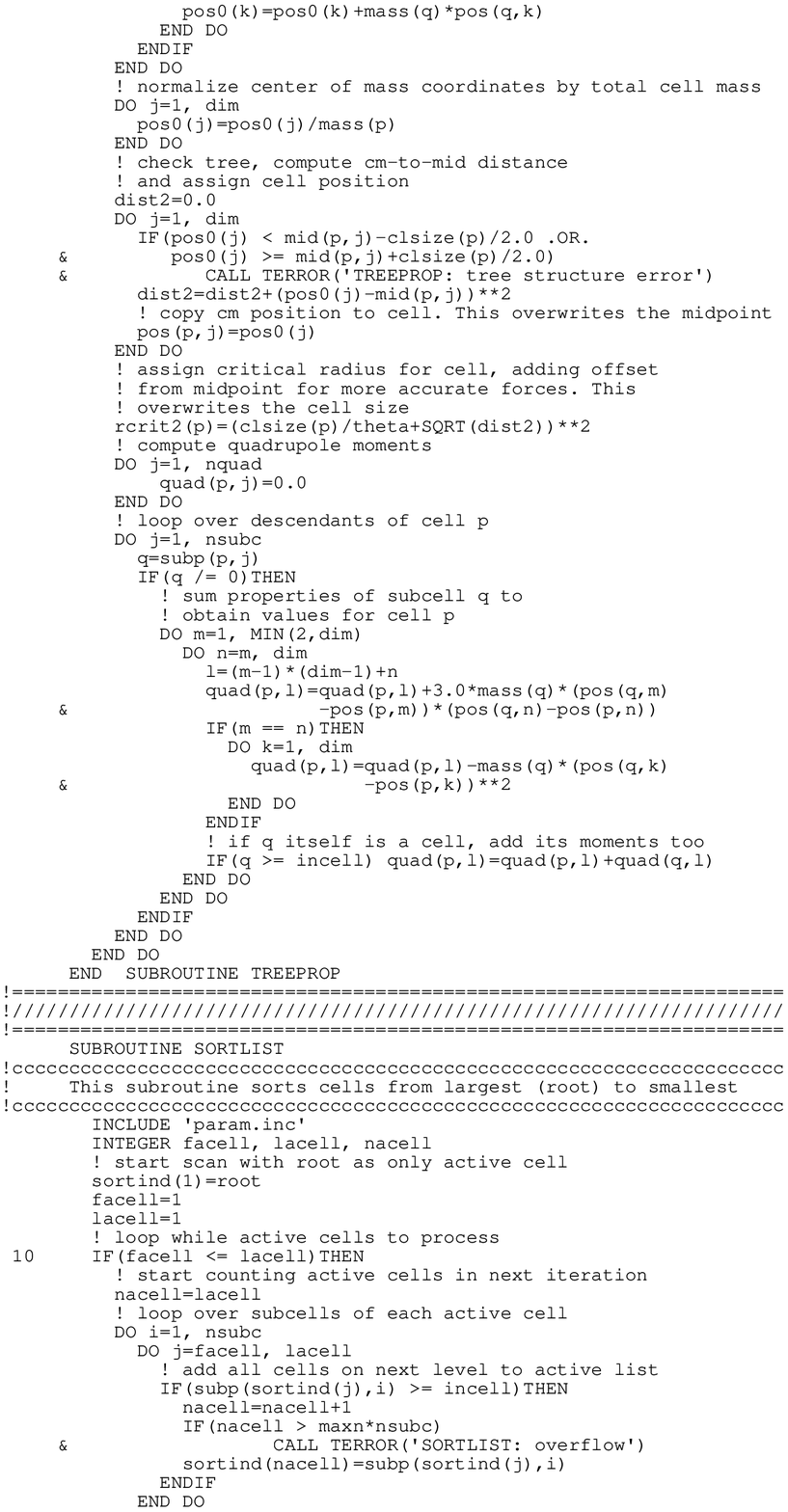}
\end{figure}
\clearpage
\begin{figure}
\epsscale{0.85} \plotone{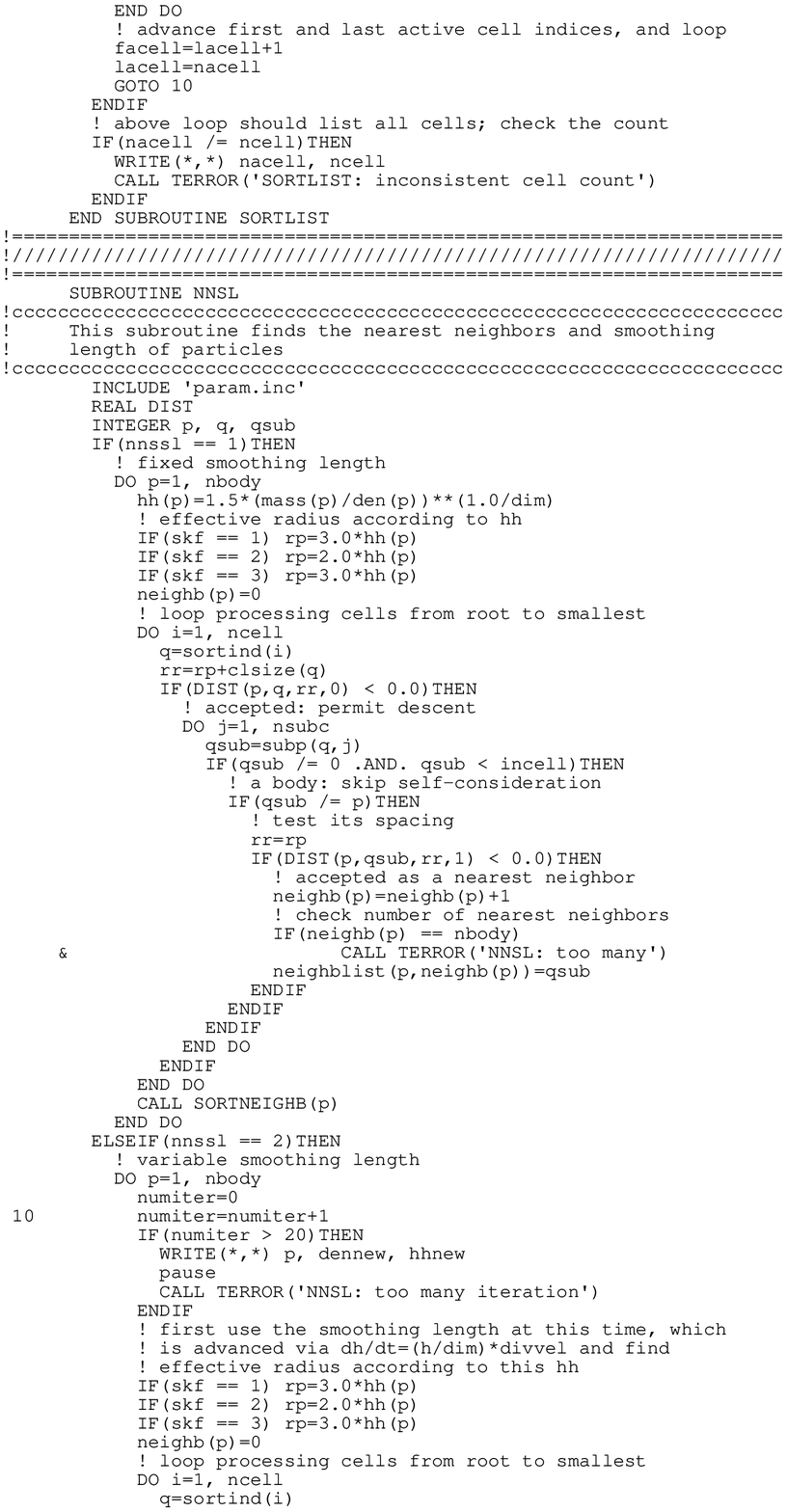}
\end{figure}
\clearpage
\begin{figure}
\epsscale{0.85} \plotone{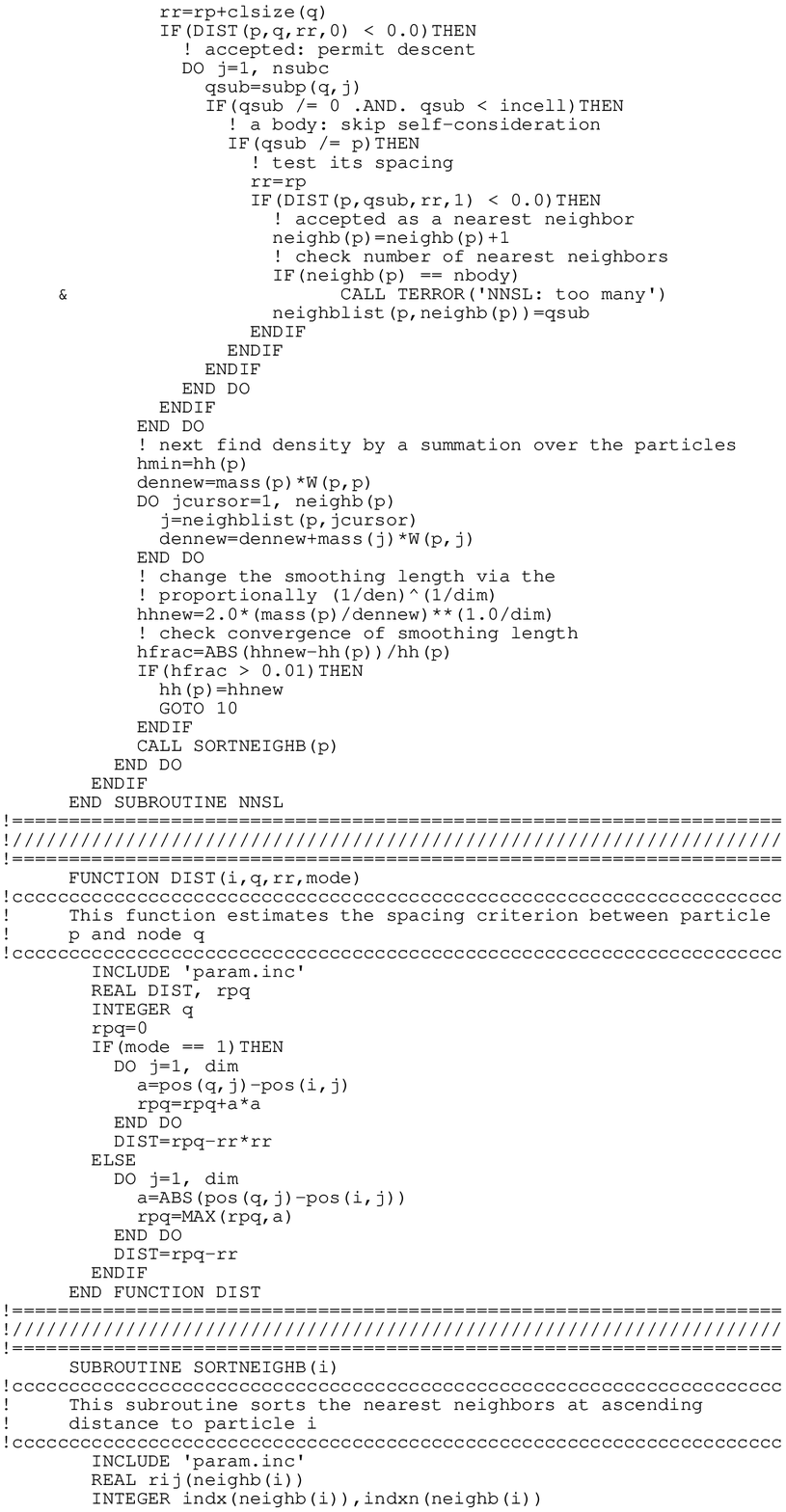}
\end{figure}
\clearpage
\begin{figure}
\epsscale{0.85} \plotone{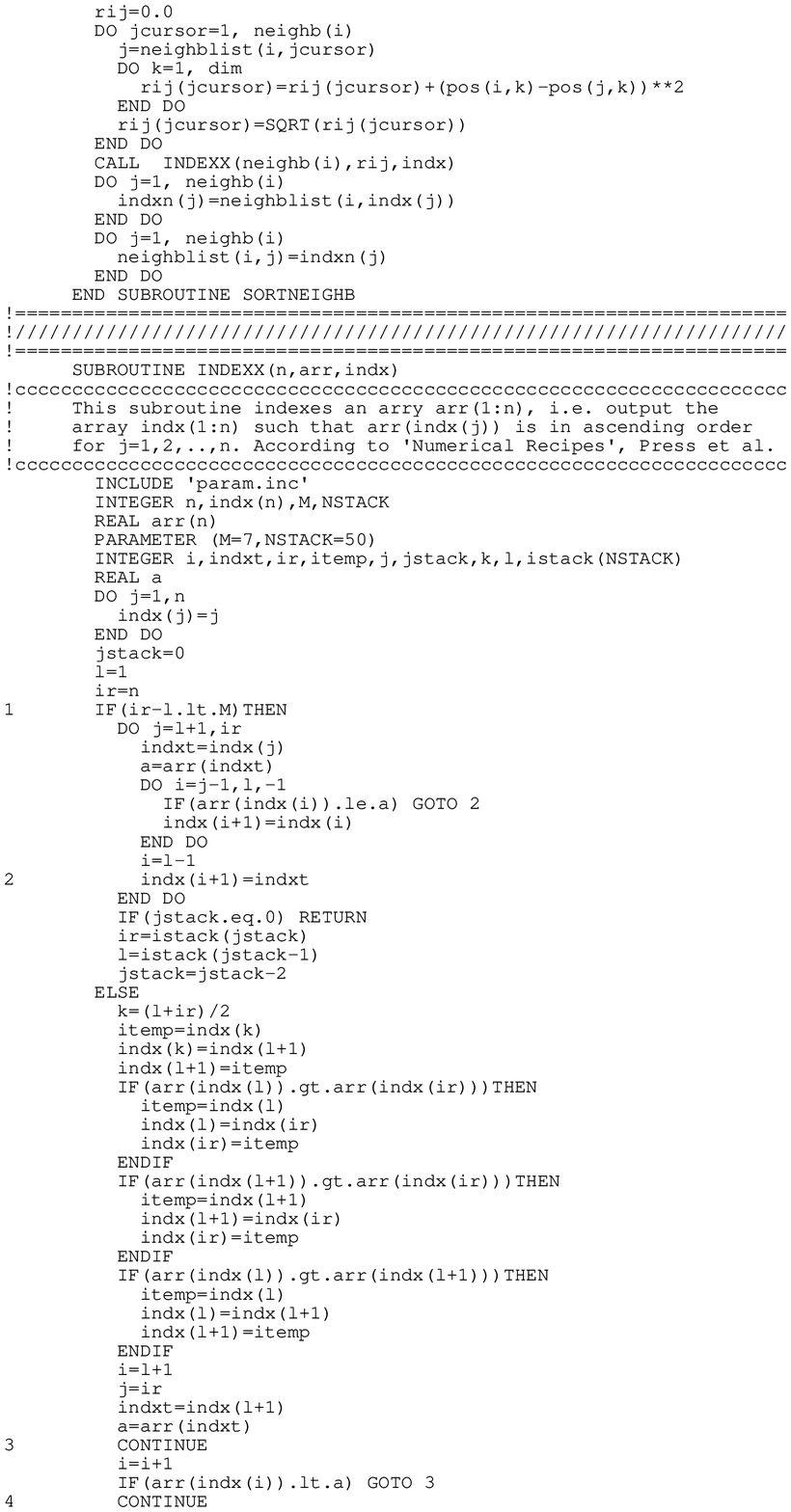}
\end{figure}
\clearpage
\begin{figure}
\epsscale{0.85} \plotone{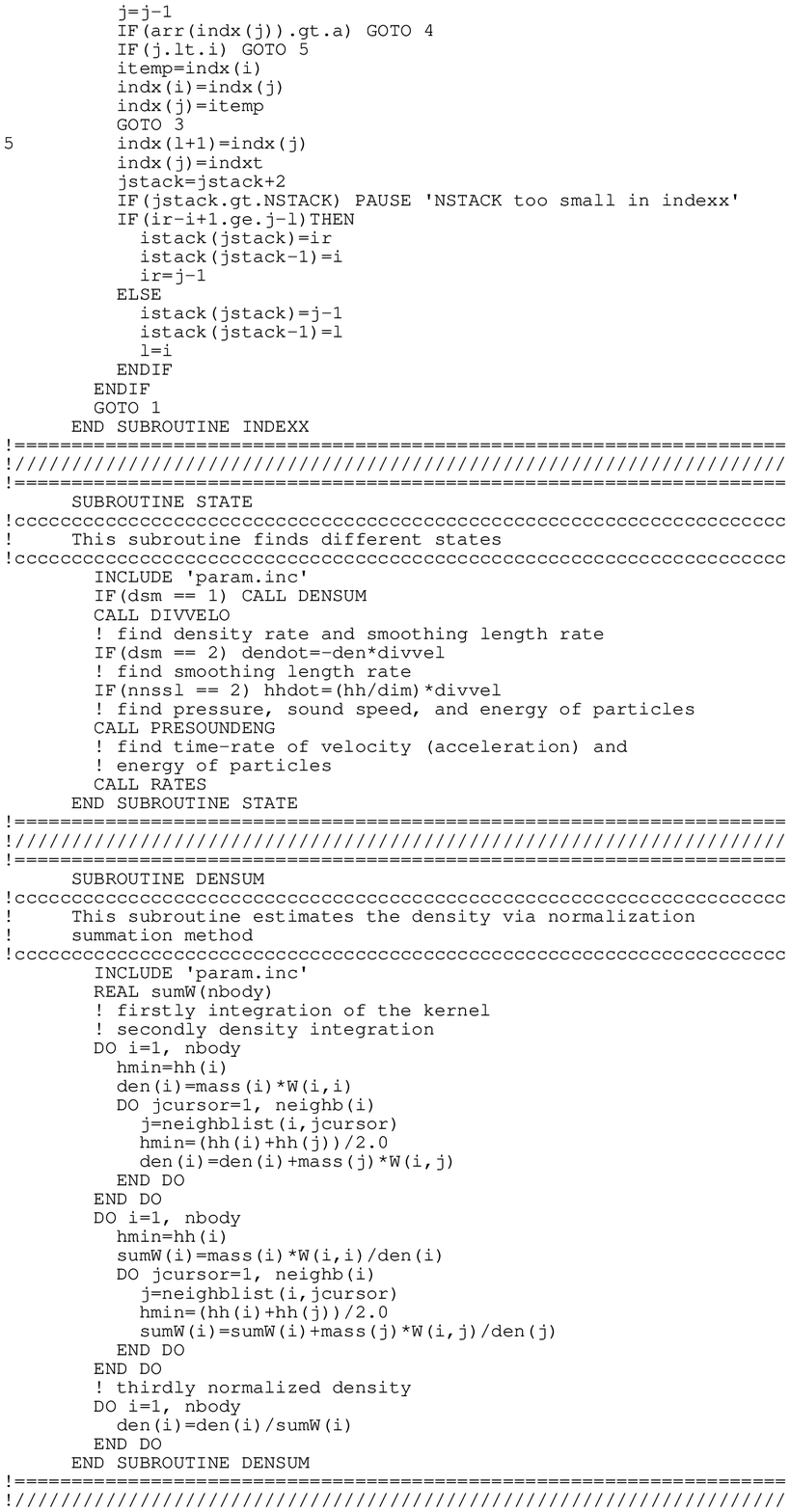}
\end{figure}
\clearpage
\begin{figure}
\epsscale{0.85} \plotone{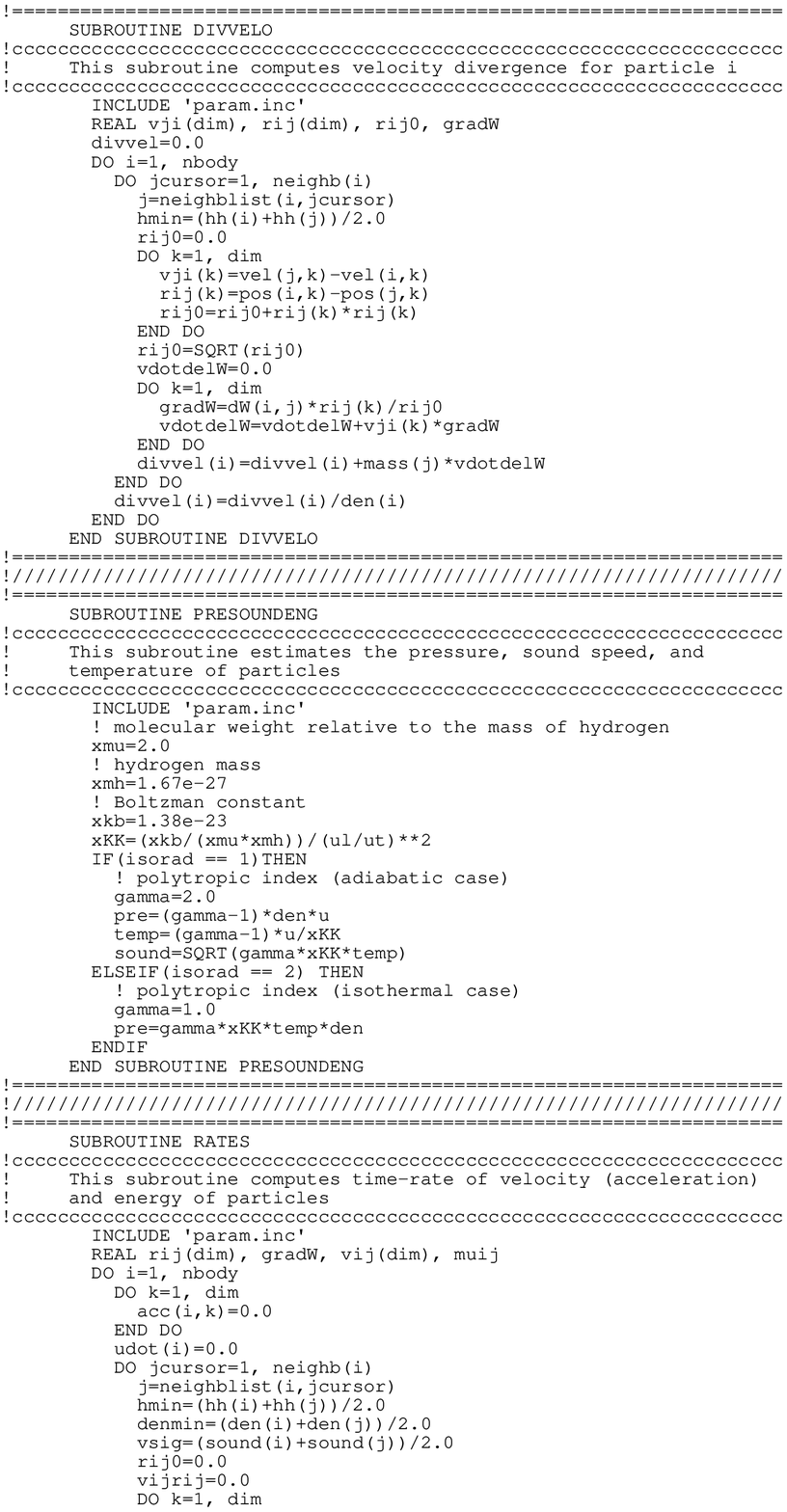}
\end{figure}
\clearpage
\begin{figure}
\epsscale{0.85} \plotone{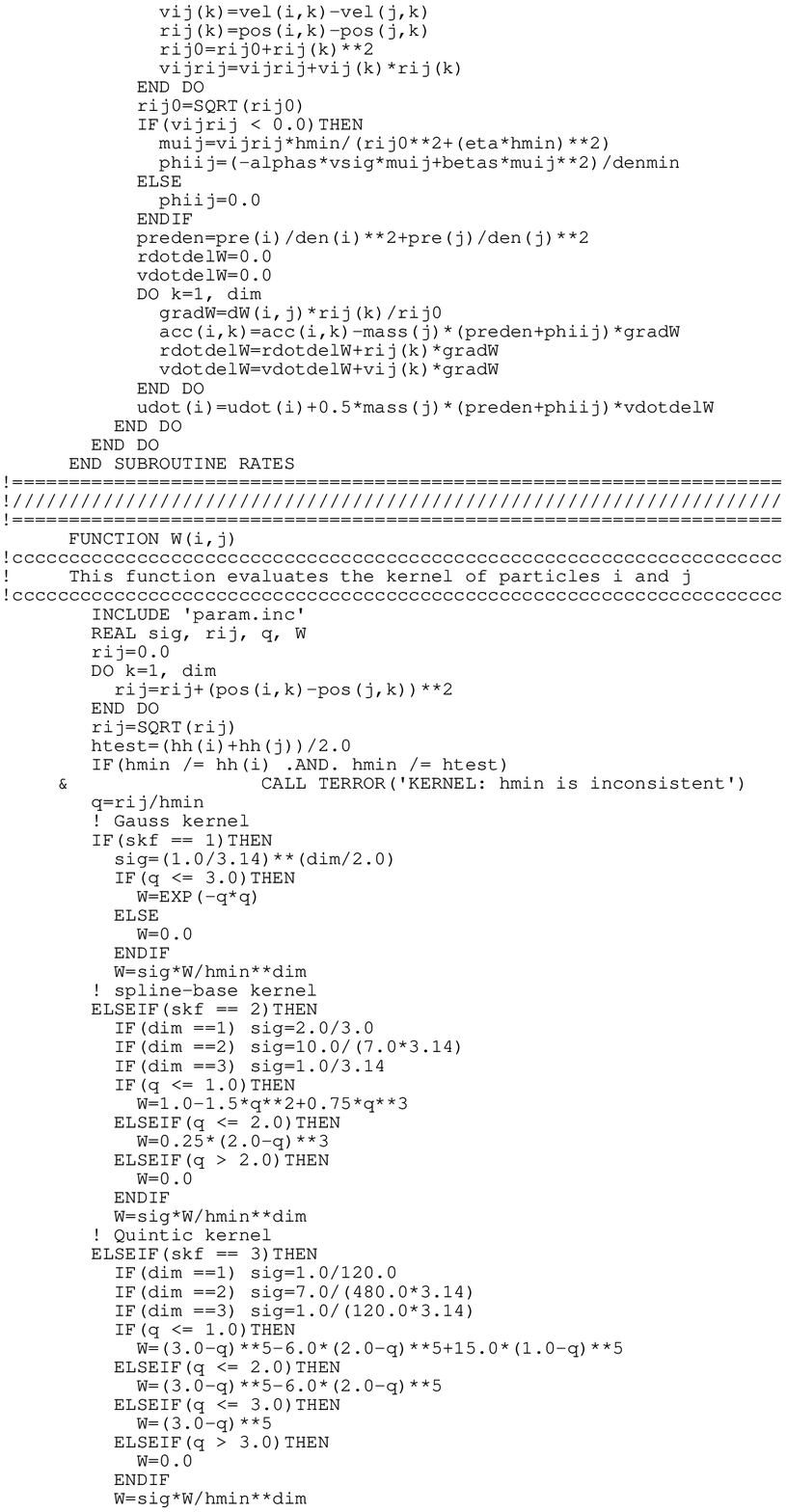}
\end{figure}
\clearpage
\begin{figure}
\epsscale{0.85} \plotone{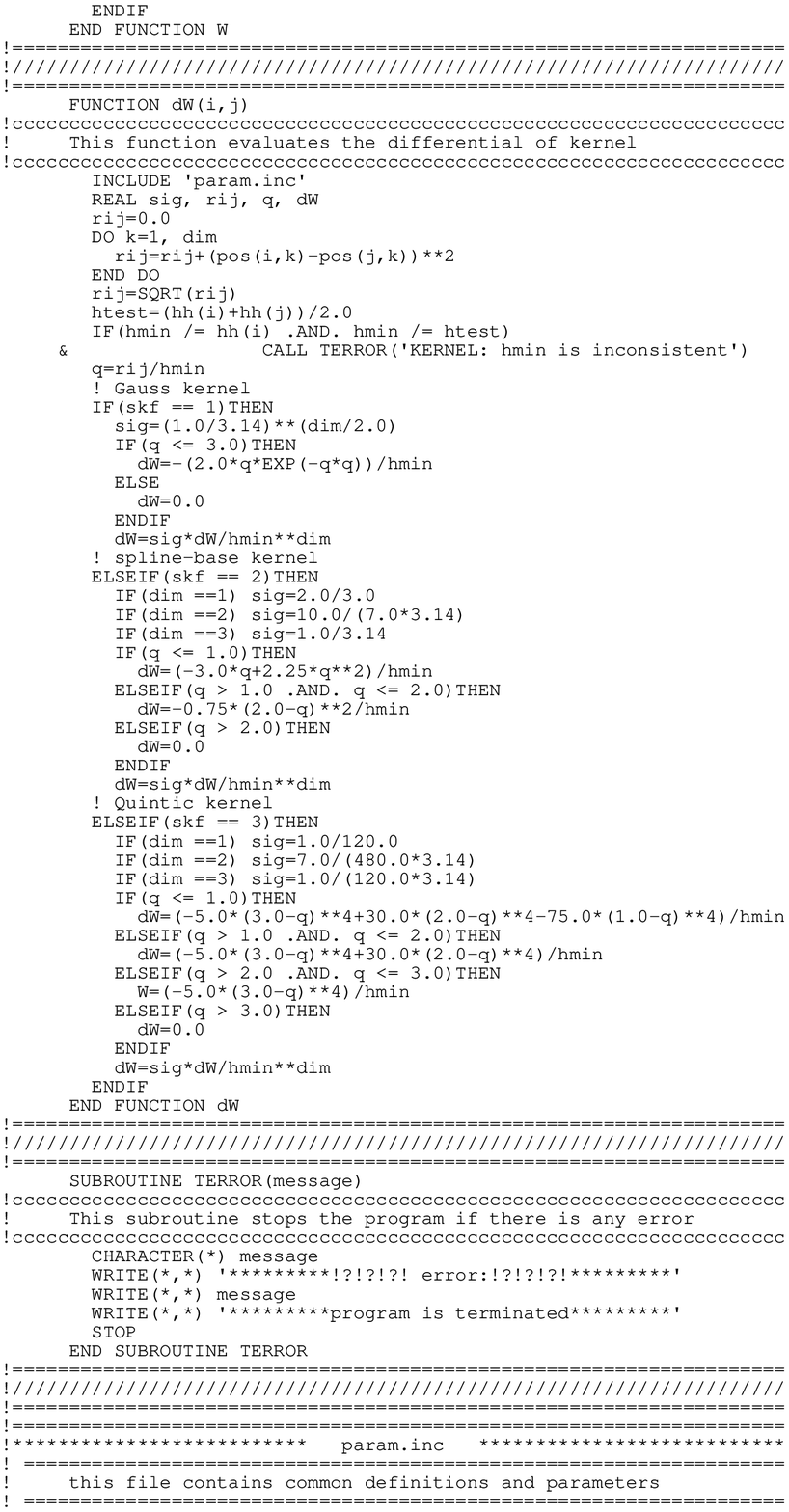}
\end{figure}
\clearpage
\begin{figure}
\epsscale{0.85} \plotone{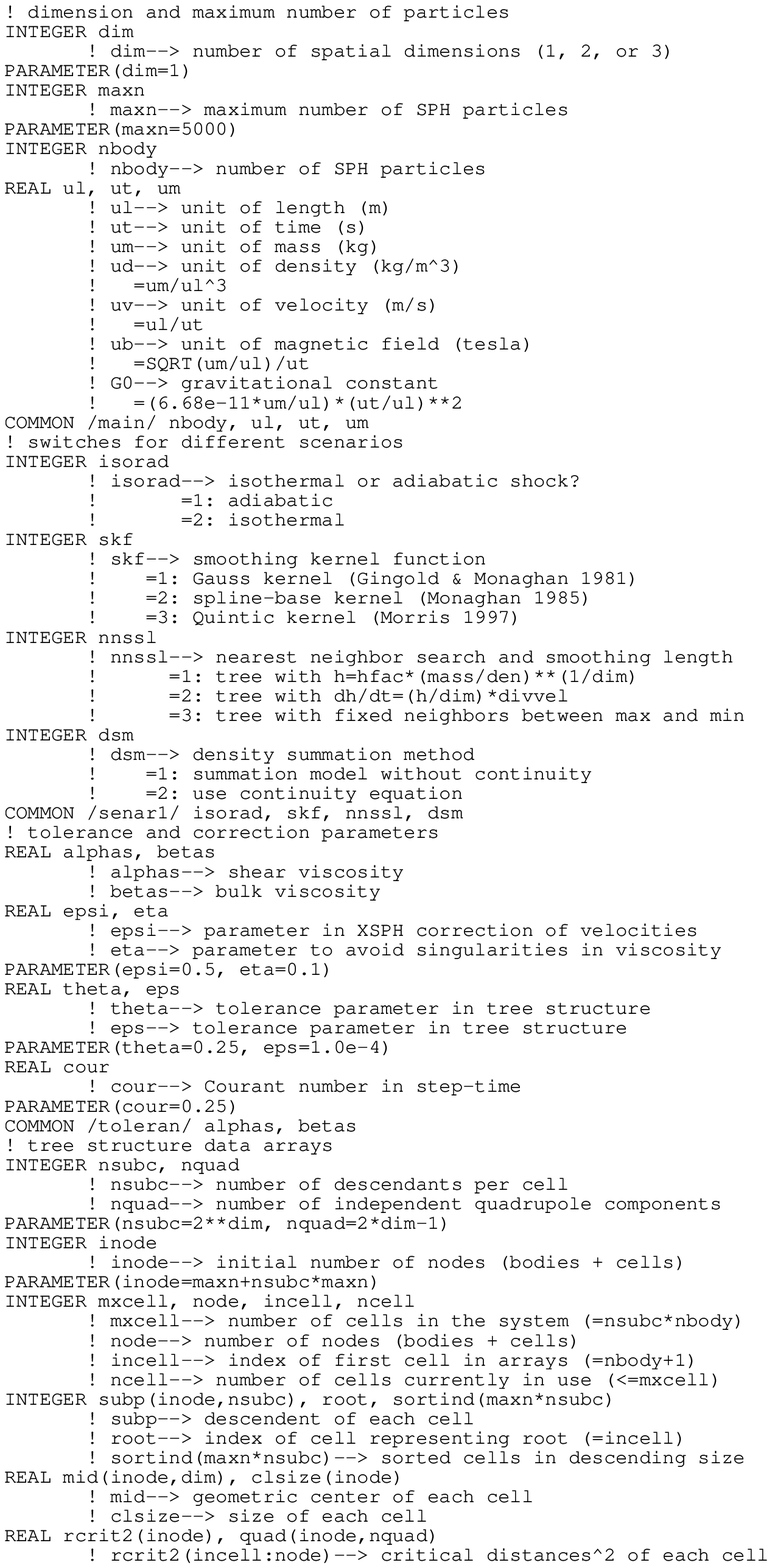}
\end{figure}
\clearpage
\begin{figure}
\epsscale{0.85} \plotone{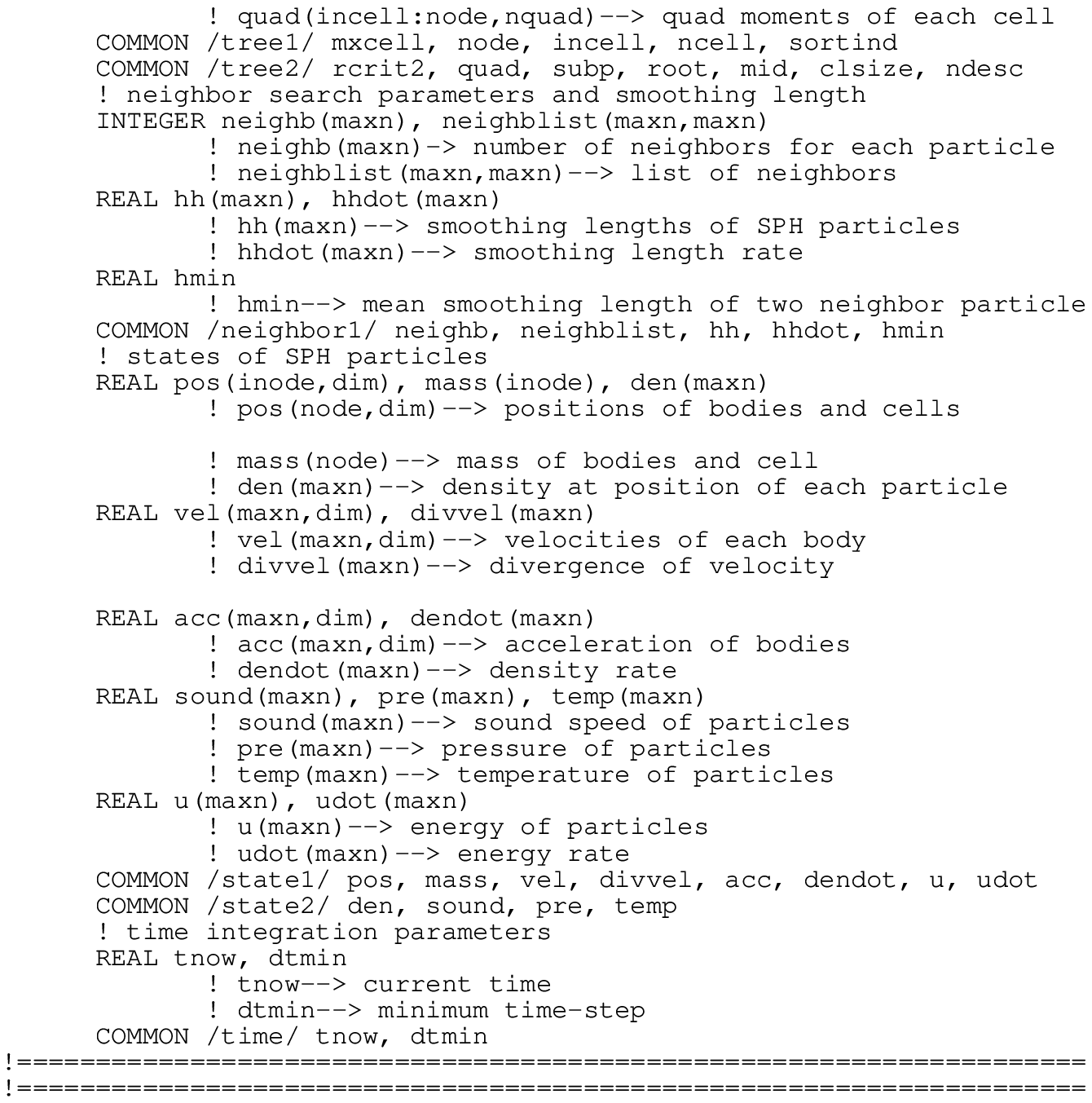}
\end{figure}

\end{document}